\documentclass[manuscript,12pt]{aastex631}

\shorttitle{Asteroids \& $J_2$}
\shortauthors{Zeebe and Kocken}

\graphicspath{{./}{fig/}}

\newcommand{\orbN}{{\tt orbitN}}
\newcommand{\hnb}{{\tt HNBody}}
\newcommand{\JT}{\mbox{$J_2$}}
\newcommand{\mA}{\mbox{$m_A$}}
\newcommand{\NA}{\mbox{$N_A$}}
\newcommand{\fmA}{\mbox{$fm_A$}}

\newcommand{\gftL}{\mbox{($g_4$$-$$g_3$)}}

\newcommand{\sftL}{\mbox{($s_4$$-$$s_3$)}}
\newcommand{\sftA}{\mbox{$|s_4$$-$$s_3|$}}
\newcommand{\gt}{\mbox{$g_3$}}
\newcommand{\gf}{\mbox{$g_4$}}
\newcommand{\st}{\mbox{$s_3$}}
\renewcommand{\sf}{\mbox{$s_4$}}
\newcommand{\zs}{\mbox{$z^*$}}
\newcommand{\zps}{\mbox{$z'^*$}}
\newcommand{\zpps}{\mbox{$z''^*$}}
\newcommand{\zts}{\mbox{$\zeta^*$}}

\newcommand{\vpi}{\varpi}
\newcommand{\om}{\mbox{$\omega$}}
\newcommand{\Om}{\mbox{$\Omega$}}
\newcommand{\alp}{\mbox{$\alpha$}}
\newcommand{\tht}{\mbox{$\theta$}}
\newcommand{\Dtht}{\mbox{$\Delta\theta$}}
\newcommand{\zt}{\mbox{$\zeta$}}
\newcommand{\del}{\mbox{$\delta$}}
\newcommand{\D}{\mbox{$\Delta$}}
\newcommand{\atant}{\mbox{atan2}}
\renewcommand{\i}{\mbox{$\hat{\imath}$}}
\newcommand{\asy}{\mbox{$''$y$^{-1}$}}

\newcommand{\masc}{\mbox{mas~c$^{-1}$}}
\newcommand{\MS}{\mbox{$M_{\Sun}$}}
\newcommand{\bz}{\mbox{$b^{(0)}_{3/2}$}}
\newcommand{\bo}{\mbox{$b^{(1)}_{3/2}$}}

\newcommand{\e}[1]{\mbox{$\times 10^{#1}$}}
\newcommand{\x}{\mbox{$\times$}}
\newcommand{\q}{\frac}
\newcommand{\sm}{\mbox{$\sim$}}

\newcommand{\ZBETa}{\mbox{\texttt{ZB18a}}}

\newcommand{\ZBXXa}{\mbox{\texttt{ZB20a}}}

\newcommand{\myurl}{\url{www2.hawaii.edu/~zeebe/Astro.html}}
\newcommand{\npurl}{\url{www.ncdc.noaa.gov/paleo/study/26970}}
\newcommand{\giturl}{\url{github.com/rezeebe/orbitN}}
\newcommand{\zenurl}{\url{doi.org/10.5281/zenodo.8021040}}

\begin{document}

\title{Reduced solar quadrupole moment compensates for lack of
asteroids in long-term solar system integrations}

\email{zeebe@soest.hawaii.edu}

\author[0000-0003-0806-8387]{Richard E. Zeebe}
\author[0000-0003-2196-8718]{Ilja J. Kocken}
\affiliation{
SOEST, University of Hawaii at Manoa, 
1000 Pope Road, MSB 629, Honolulu, HI 96822, USA. \\ \\
{{\rm Final Accepted Version} \\
The Astronomical Journal} \\
\rm \today
}

\begin{abstract}
State-of-the-art long-term solar system integrations
include several second order effects
such as the Sun's quadrupole moment $J_2$ and 
a contribution from asteroids (plus the Moon and general 
relativity). We recently showed that including 10 asteroids 
and a reduced $J_2$ in our astronomical solutions provides 
the best match with geologic data to $-58$~Myr.
However, the rationale for the reduced $J_2$ remained
ambiguous and may suggest that parameters for 
long-term integrations compatible with geologic observations
are not fully compatible with our knowledge of the current 
solar system (specifically $J_2$). Here we show that a reduced 
$J_2$ compensates for a diminished asteroid population in long-term 
solar system integrations, which may appear surprising. 
We present an analysis and offer a mechanism for the 
long-term compensating effects of $J_2$ and asteroid mass
in the solar system (not planetary systems in general).
Our analysis suggests that ``differential effects'' on 
specific secular frequencies involved in resonant terms 
(i.e., \gftL\ and \sftL), are critical in the long term, 
rather than short-term effects on the orbital elements of 
individual planetary orbits across the board.
Also, our results indicate that if long-term 
intergrations including the full asteroid population were 
computationally feasible, a $J_2$ value (within errors) compatible 
with our current knowledge of the solar system could be used.
Attempts to improve the long-term accuracy of astronomical 
solutions by, e.g., tinkering with initial conditions using 
current/future astronomical observations are futile 
unless asteroid deficiencies in the solar system model are 
addressed.
\end{abstract}

\section{Introduction \label{sec:intro}}

Reliable studies of the long-term dynamics of planetary systems 
require numerical integrations that are accurate and fast.
The most up-to-date long-term astronomical solutions for the 
solar system have been obtained using symplectic integrators
\citep[e.g.,][]{wisdom91,yoshida90}
that advance the equations of motion for the main solar system bodies.
In addition, several second order effects need to be considered
for accurate applications,
such as the Sun's quadrupole moment \JT\ and a contribution 
from asteroids (as well as the Moon and general relativity).
While the effects from \JT\ and asteroids
may appear negligible, their contributions become critical
for astronomical solutions over, e.g., 50-Myr time scale due 
to chaos \citep{laskar11ast,zeebe17aj}. 
Note that the focus of the present study is the past 
50-100~Myr over which the asteroid belt may be considered
essentially stable \citep[as opposed to changes on Gyr-time
scale, e.g.,][]{ito06ast,minton10,morbidelli14,izidoro15,levison15}.
The limitations imposed by dynamical chaos are fundamental and of
physical nature, rather than numerical, and severely limit
our understanding and ability to reconstruct and predict the 
solar system's history and long-term future \citep{abbot21,
zeebelantink24aj}. The chaos also imposes strong limits on 
widely used geological and astrochronological 
applications such as developing a fully calibrated astronomical 
time scale beyond \sm{50}~Ma \citep[for recent efforts to 
tackle the issue, see][]{zeebelourens19,zeebelourens22epsl,
kockenzeebe24pa}. 
For example, \citet{zeebelourens19} used geological data
to constrain and select suitable astronomical solutions, thereby 
extending the fully calibrated astronomical time scale to $-58$~Myr 
and providing highly accurate absolute ages and chronologies.

Accurate numerical solar system integrations over tens of millions 
of years are still computationally expensive and CPU times for $N$
fully interacting bodies generally scale with $N^2$. As a result, 
only a relatively small number of asteroids (\NA) are usually
included in long-term integrations. \citet{zeebelourens19},
for instance, included 10 asteroids in their
astronomical solution \ZBETa. Surprisingly,
the best match with geologic data was obtained with a reduced 
$\JT \simeq 1.3 \e{-7}$ (\ZBETa), compared to reported modern 
values of \sm{1.8} to $\sm{2.3}\e{-7}$ \citep[for summary, see e.g.,]
[]{rozelot20,zwaard22,alves25}. Subsequently, \citet{zeebelourens22epsl}
showed that including 50 asteroids and setting $\JT = 1.47\e{-7}$ 
yielded a solution (\ZBXXa) practically indistinguishable from \ZBETa\ 
to $-58$~Myr, suggesting a dynamical link between \JT\ and \NA.  
Note that for \ZBXXa, 10 asteroids were included as 
heavyweight particles (HWPs) and 40 as lightweight particles (LWPs).
HWPs are subject to the same full interactions as the planets,
while LWPs are dynamically equivalent to HWPs but self-gravity 
(LWP-LWP forces) is ignored; HWP-LWP interactions are included 
\citep[see][]{rauch02}.
Hitherto, the cause and rationale for the reduced \JT\ (relative
to modern values) and the link between \JT\ and \NA\
remained somewhat ambiguous and may suggest that parameters 
required for long-term integrations compatible with geologic 
observations of the past are not fully compatible with our 
knowledge of the current solar system (specifically \JT). 
In the present study, we show that a reduced \JT\ compensates 
for a diminished asteroid population \NA\ (or total mass \mA) 
in long-term solar system integrations ($\mathcal{O}(10^8$~y)). 
We also present an analysis of our numerical integration results 
that offers a mechanism for the long-term opposing effects of 
\JT\ and \NA\ (or \mA).

The results of our study are important for testing the credibility
of our current solar system models, their numerical long-term 
integration, as well as the compatibility of our astronomical 
solutions with geologic data.
Also, because \JT\ and asteroids have a significant
impact on the long-term behavior of astronomical solutions,
our findings underline that
attempts to improve their accuracy by,
e.g., tinkering with initial conditions using current/future 
astronomical observations are futile unless asteroid 
deficiencies in the solar system model are addressed.

\subsection{Complexity of Chaotic Dynamics \label{sec:coc}}
The long-term dynamics of the full solar system are
complex and, due to chaotic behavior, not necessarily allow 
simple explanations and straightforward mechanisms. Solar system 
chaos has been studied
for over thirty years, yet the large-scale consequences
of the chaos and the resonances involved remain an area of 
active research \citep[e.g.,][]{batygin15,zeebe17aj,abbot21,
hernandez22,mogavero22,zeebelantink24aj}. 
For instance, long-term integrations
show that a difference of a few centimeters in the initial 
position of Mercury may in one case lead to a stable system 
over 5~Gyr, but to ejection of Mercury from the 
solar system in another. This behavior is ascribed
to ``{\it sensitivity to initial conditions}'',
i.e., the exponential divergence of trajectories.
Also, the dynamics during the final stages of instability
are comprehensible, once a certain resonance has been 
excited. However, there is no simple physical mechanism that 
allows following the dynamics across all time scales and
allows predicting, for example, which initial conditions 
lead to stability and instability. The present study is 
concerned with
small perturbations due to, e.g., asteroidal mass, which 
is of order $10^{-10}$ relative to the solar mass. The 
compensating effect relative to \JT\ only plays out over 
tens of millions of years and involves the resonance 
$m\gftL - \sftL$, where $m = 1,2$ (see 
Section~\ref{sec:anyss}). Thus, we are dealing with initially 
small effects, gradual changes over time, and chaotic 
dynamics. We present an analysis and offer a 
mechanism for the long-term compensating effect
of asteroidal mass and $J_2$ (Section~\ref{sec:anyss}).
However, we caution that the effect is not straightforward 
and does not appear to have a simple mechanical analogy,
for example.

\section{Low-Order Theoretical Guesses \label{sec:guess}}

To obtain a first guess of the magnitude and direction of 
the effects of \JT\ and asteroids on the orbital elements 
of a single body, we start with low-order theoretical guesses
based on equations derived using perturbation theory
available in the literature.

\subsection{Quadrupole Moment $J_2$}

The secular effect of the Sun's gravitational quadrupole moment 
on the orbital 
elements of a single body to order \JT\ may be estimated 
as \citep[in milliarcsec per centrury, see e.g,][]{danby88}:
\begin{eqnarray}
\dot{\om}  & = & - \q{3 n J_2 R^2_S}{2a^2 (1-e^2)^2} \ 
                   (5/2 \ \sin^2{I} - 2)  
             =      1.81~\masc  \label{eqn:omSJ} \\
\dot{\Om}  & = & - \q{3 n J_2 R^2_S}{2a^2 (1-e^2)^2} \ \cos(I)
             =     -0.918~\masc \label{eqn:OmLJ} \\[2ex]
\dot{\vpi} & = &  \dot{\om} + \dot{\Om} = 0.897~\masc \ ,
\end{eqnarray}
where $\om$ is the argument of perihelion of the body's orbit,
$n$ the mean motion,
$a$ the semimajor axis,
$e$ the eccentricity,
$I$ the inclination,
$\Om$ the longitude of the ascending node,
$\vpi$ the longitude of perihelion, 
$R_S = 0.00465$~au is the solar radius, and
$\JT = 2.2 \e{-7}$ the solar quadrupole moment.
The values inserted above refer to Earth's present orbit, i.e., 
$a = 1$~au, $e = 0.0167$, $I = 7.155^\circ$ 
\citep[HCI frame,][]{zeebe17aj}.
Thus, in this simple view, 
\JT\ causes the line of apsides to advance and the line of 
nodes to regress.
For comparison, numerical integrations with \orbN\ 
\citep{zeebe23aj} over 1,000 years (only Sun + Earth)
yields linear trends in \om\ and \Om, with essentially the 
same results, i.e.,
$\dot{\om} =  1.82 $~\masc\ and
$\dot{\Om} = -0.918$~\masc.

\subsection{Asteroid}

The secular effect of a single asteroid on the orbital elements 
of a single body may be estimated as \citep[see][who derived an equation 
for $\dot{\vpi}$ rather than $\dot{\om}$]{kuchynka10}:
\begin{eqnarray}
\dot{\om}  & = &  \dot{\vpi} - \dot{\Om} = 3.32~\masc 
                  \label{eqn:omSm} \\
\dot{\Om}  & = & - n \alp^2 \q{m_A}{4 \MS} \bo   
     \left[
         1 - \q{I_A}{I} \cos (\Om - \Om_A)
    \right] = -4.16~\masc  \label{eqn:OmLm} \\[2ex]
\dot{\vpi} & = & n \alp^2 \q{m_A}{4 \MS} 
     \left[
         \bo + \q{e_A}{e} 
             \left(
                3 \bz - 2 \bo (\alp + \alp^{-1})   
             \right) \cos (\vpi - \vpi_A)
    \right] \nonumber \\
          & = & -0.845~\masc \label{eqn:vpim} \ ,
\end{eqnarray}
where the subscript `$A$' refers to the asteroid,
$\alp = a/a_A < 1$, \MS~is the mass of the central body,
and~\bz\ and~\bo\ are Laplace coefficients that depend 
on $\alp$. 
The values above refer to Earth's and Ceres' present orbits.
In this simple view, the asteroid {\sl also} causes the line of 
apsides to advance and the line of nodes to regress.
Note that theoretical estimates for a population of
asteroids, or an asteroid ring, yield very similar
results, primarily depending on the assumed mass 
and semimajor axis \citep{krasinsky02,kuchynka10,pitjeva13}.
For comparison, numerical integrations with \orbN\ 
over 1,000 years (only Sun + Earth + Ceres) yields linear
trends in \om\ and \Om, with
$\dot{\om} =  3.53 $~\masc\ and
$\dot{\Om} = -4.07$~\masc.

\subsection{Intermediate Time-Scale Guesses}

The secular, low-order theoretical guesses given above refer 
to present solar system orbital parameters, which do however
change over time. Guesses on intermediate time scale may 
be obtained by inserting orbital parameters as calculated 
by our full solar system integrations into the equations above
(say, over a few million years). 
The secular effect of \JT\ for changing $e$ and $I$ is
straightforward (see Eqs.~(\ref{eqn:omSJ}) and (\ref{eqn:OmLJ})).
Given that $e$ and $I$ and their changes are small,
the variations in $\dot{\om}$ and $\dot{\Om}$ around
the values given above are small. For example, estimated
mean values over the past 2~Myr are
$[\dot{\om},\ \dot{\Om}] = [1.8277,\ -0.9215]$~\asy.
Estimating the secular effect of a single asteroid is 
slightly more complex because of the terms involving the
angles $\Om$ and $\vpi$ (see Eqs.~(\ref{eqn:OmLm}) and 
(\ref{eqn:vpim})). Estimated mean values over the past 
2~Myr are
$[\dot{\om},\ \dot{\Om}] = [2.6528,\ -0.0126]$~\asy.
Nevertheless, the signs obtained for changes in $\dot{\om}$ 
and $\dot{\Om}$ on intermediate time scale are the same
as for present orbital parameters (Eqs.~(\ref{eqn:omSJ}),
(\ref{eqn:OmLJ}), (\ref{eqn:omSm}), and (\ref{eqn:OmLm})).

\subsection{Combined Effects}

Importantly, the theoretical estimates above suggest
that \JT\ and asteroids have the same
qualitative effect on \om\ and \Om, i.e., positive
for $\dot{\om}$ and negative for $\dot{\Om}$.
As a result, a simultaneous increase in \JT\ and asteroid 
mass would be expected to enhance each other.
However, an enhancement is opposite to the 
results of our long-term integrations
(see Section~\ref{sec:num}), which show that an increased 
$J_2$ and a larger asteroid mass tend to cancel each 
other out in the long run (\sm{50}~Myr).
In summary, the low-order theoretical effects of
\JT\ and asteroid mass on the orbital elements of a 
single body do not explain the dynamics revealed in our 
long-term numerical integrations of the solar system. 
Additional analyses are required to understand the
long-term dynamics (see Section~\ref{sec:mech}).

Note that similar short-term trends over 1,000 years
as derived above 
($\dot{\om} > 0, \dot{\Om} < 0$) were also confirmed 
by our full solar system integrations (see 
Section~\ref{sec:num}). Only in that case, we compared 
{\sl differences} ($\D$'s) for individual planetary orbits 
between runs with separate increases in \JT\ and \mA.
For all inner planets, we found short-term 
$(\dot{\D \om}) > 0$ and $(\dot{\D \Om}) < 0$ for 
both \JT\ and \mA\ increases.

\section{Numerical Integrations \label{sec:num}}

To study the long-term dynamical link between \JT\ and 
asteroids in orbital solutions, we performed a number of 
numerical experiments based on full solar system 
integrations.

\subsection{Numerical Methods}

For the long-term solar system integrations, we tested
the integrator packages \orbN\ \citep{zeebe23aja} 
and \texttt{HNBody} \citep{rauch02}.
For consistency with the original \ZBETa\ solution,
we used \texttt{HNBody 1.0.10} with the \texttt{lunar} 
module and integrated the solar system from $t = 0$ to
$-100$~Myr \citep[for details see, Appendix~\ref{sec:bset} 
and][]{zeebe17aj,zeebelourens19}.
We ran two sets of 128 parameter variations in parallel
on the high performance compute cluster Derecho 
\citep{cisl23}.
Set~1 and~2 included 100 and 10 asteroids, respectively.
In addition, Set~2 included an artificial, uniform mass 
increase of each asteroid.
For the base runs, we started with slightly higher parameter 
values (\JT,$\x \fmA$) than in \ZBETa\ (1.3050,$\x 1.0$), 
i.e., (1.40,$\x 1.0$) 
for Set~1 and (1.40,$\x 1.2$) for Set~2, where $\JT \x 10^7$ 
and $\fmA$ is a factor describing the mass increase in \mA\
(see Table~\ref{tab:exp}).

\begin{table}[h]
    \centering
    \caption{
    Experimental design of full solar system integrations. 
    } 
    \begin{tabular}{lcrlcc}
    \tableline\tableline
    Label & \JT\e{7} & Asteroids & \fmA & No. & Total\\
    \tableline
    A100a & 1.4--1.78750 & 1\textsuperscript{st} 90 & \x 1 & 32 & \\
    A100b & 1.4--1.78750 & 2\textsuperscript{nd} 90 & \x 1 & 32 & \\
    A100c & 1.4--1.78750 & 3\textsuperscript{rd} 90 & \x 1 & 32 & \\
    A100d & 1.4--1.78750 & massive 90 & \x 1 & 32 & 128\\
    \tableline
    A10a  & 1.4--1.98125 & $^a$ `named' 10 & \x 1.2 & 32 & \\ %
    A10b  & 1.4--1.98125 & `named' 10 & \x 1.4 & 32 & \\
    A10c  & 1.4--1.98125 & `named' 10 & \x 1.6 & 32 & \\
    A10d  & 1.4--1.98125 & `named' 10 & \x 1.8 & 32 & 128\\
    \hline
    \end{tabular}
    \label{tab:exp}

\noindent {\small
$^a$For `named' 10 asteroids, see Table~\ref{tab:ast}. \\[0ex]
}
\end{table}

\subsubsection{Number of Asteroids \label{sec:nast}}

Initial tests indicated that increasing the total number 
of asteroids in the simulation to 300 would require 
unrealistic computing demands, i.e., wallclock times up to 
2500~hours ($>$100~days) for a 100~Myr simulation with 
either integrator \texttt{orbitN} and \texttt{HNBody} 
using only HWPs. Even a mix of HWPs and LWPs using
HNBody would require $\sm{600}$~hours ($>$25 days).
Thus, we ran four groups of simulations with 100 asteroids total
(Set~1), where the planets, Pluto, and 10 ``named asteroids''
were treated as HWPs (see Table~\ref{tab:ast} and Appendix~\ref{sec:bset}).
The remaining 90 ``unnamed asteroids'' were treated as 
LWPs (wallclock time of $\sm 200$ hours). The unnamed 
asteroids were taken from a list of 300 asteroids ordered by 
semimajor axis as included in the DE430/431 ephimerides 
\citep{folkner14}. The four groups of asteroids consisted 
of the 10 named asteroids and different selections of 
unnamed asteroids. Group
\texttt{A100a} included the first 90, 
\texttt{A100b} included the second 90, and 
\texttt{A100c} included the third 90 asteroids of the list of 300, 
sorted by semi-major axis; group
\texttt{A100d} included the 90 most massive asteroids
(Table~\ref{tab:exp}).
We varied $\JT \x 10^7$ between $1.4$ and $1.7875$
in increments of $0.0125$, resulting in 128 new
astronomical solutions.
To compare the new astronomical solutions to the reference
solution \ZBETa, we scaled
Earth's eccentricity ($e_{\oplus}$) for \ZBETa\ (subtracted the 
mean and divided by the standard deviation), as well as the 
new solutions between $-58$ and $-50$~Myr and calculated the 
Root Mean Square Deviation (RMSD). 
The time interval $-58$ to $-50$~Myr was selected
because \ZBETa\ has been constrained by geological data to 
$-58$~Myr and state-of-the-art astronomical solutions agree to
$-50$~Myr (i.e., their RMSDs~$\simeq 0$ for $t > -50$~Myr).
The RMSD is given by:
\begin{equation}
 \text{RMSD} = \sqrt{\frac{1}{n}
               \sum_{i=1}^{n}(e_{i} - \hat{e}_{i})^{2}} \ ,
\label{eqn:rmsd}    
\end{equation}
where $e$ and $\hat{e}$ are the scaled eccentricity 
of the new and reference solution, respectively, 
and $n$ is the total number of values (output steps within
the considered time interval).

\def\tx{-1ex}
\begin{table}[t]
\centering
\caption{First 10 asteroids included in all simulations as HWPs.
         \label{tab:ast}}
\begin{tabular}{rlc}
\tableline\tableline
\#       & Name       & Mass$^a$ \\
\tableline  
 1 & Vesta      & 1.30\e{-10} \\ [\tx]
 2 & Ceres      & 4.73\e{-10} \\ [\tx]
 3 & Pallas     & 1.05\e{-10} \\ [\tx]
 4 & Iris       & 7.22\e{-12} \\ [\tx]
 5 & Bamberga   & 4.69\e{-12} \\ [\tx]
 6 & Hygiea     & 4.18\e{-11} \\ [\tx]
 7 & Euphrosyne & 2.14\e{-11} \\ [\tx]
 8 & Interamnia & 1.78\e{-11} \\ [\tx]
 9 & Davida     & 1.76\e{-11} \\ [\tx]
10 & Eunomia    & 1.58\e{-11} \\ [\tx]
\tableline
\end{tabular}

\noindent {\small
$^a$In solar masses \citep{folkner14}.        \\[0ex]
}
\end{table}

\begin{figure*}[t]
    \hspace*{-08ex}
    \centering
    \includegraphics[scale=1.2]{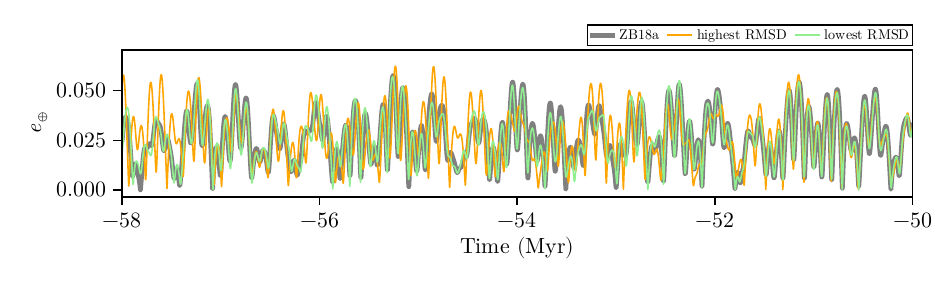}
    \caption{
    \footnotesize
    \label{fig:ecc}
    Earth's eccentricity, $e_{\oplus}$, of the best matching 
    solution (i.e., lowest RMSD~$=0.233$, \JT~$=1.8125$ 
    for the \texttt{A10d} experiment, green) matching the 
    reference \ZBETa\ solution (gray) very well between $-58$ and
    $-50$~Myr, where \ZBETa\ has been geologically constrained.
    In contrast, the worst matching solution
    (i.e., highest RMSD~$=0.95$, \JT~$=1.6625$ for 
    \texttt{A10d}, orange) provides a poor match prior to
    $-52$~Myr.
    }
\end{figure*}

\subsubsection{Asteroid Mass}

The results of the first set of experiments ($\NA = 100$) 
suggested that the total mass of the asteroids, \mA, plays 
a significant role in compensating for the effects of increased 
\JT\ values. The total mass of the 10 named asteroids 
(Table~\ref{tab:ast})
is equivalent to $\sm{8.34} \e{-10}$ 
solar masses (\MS) \citep[see DE430/431 ephimerides,][]{folkner14}.
Recent estimates for the total mass of 
the asteroid belt are $(12.04 \pm 0.09) \e{-10} \MS$, 
with a reported $3\sigma$ error for that particular
method \citep{pitjevapitjev18}. Other methods yield
estimates of up to $13.6\e{-10}\MS$
\citep[see e.g.,][]{kuchynkafolkner13,demeo13}.
Thus, an artificial mass increase by~44\% and~66\%,
respectively, would scale up the mass of the 10 asteroids to 
approximately the estimated total mass of the asteroid belt.
To study the effects of \mA, we set up a second set of 
simulations (Set~2, $4 \x 32$ runs) using the 10
named asteroids (to allow fast computation,
$\sm 11$\,hours wallclock time) 
similar to \ZBETa, but with an artificially
increased asteroid mass by a factor of $\fmA = 1.2$ 
for \texttt{A10a}, 1.4 for \texttt{A10b}, 1.6 
for \texttt{A10c}, and 1.8 for \texttt{A10d}.
For Set~2 of the simulations, we varied $\JT \e{7}$ 
between $1.4$ and $1.98125$ in increments of $0.01875$,
resulting in 128 additional runs (Table~\ref{tab:exp}).
Below, we refer to the first simulation of Set~2, i.e., 
(\JT,$\x \fmA$) = (1.40,$\x 1.2$) as our `Base' run.

\begin{figure*}
    \centering
    \includegraphics[scale=0.7]{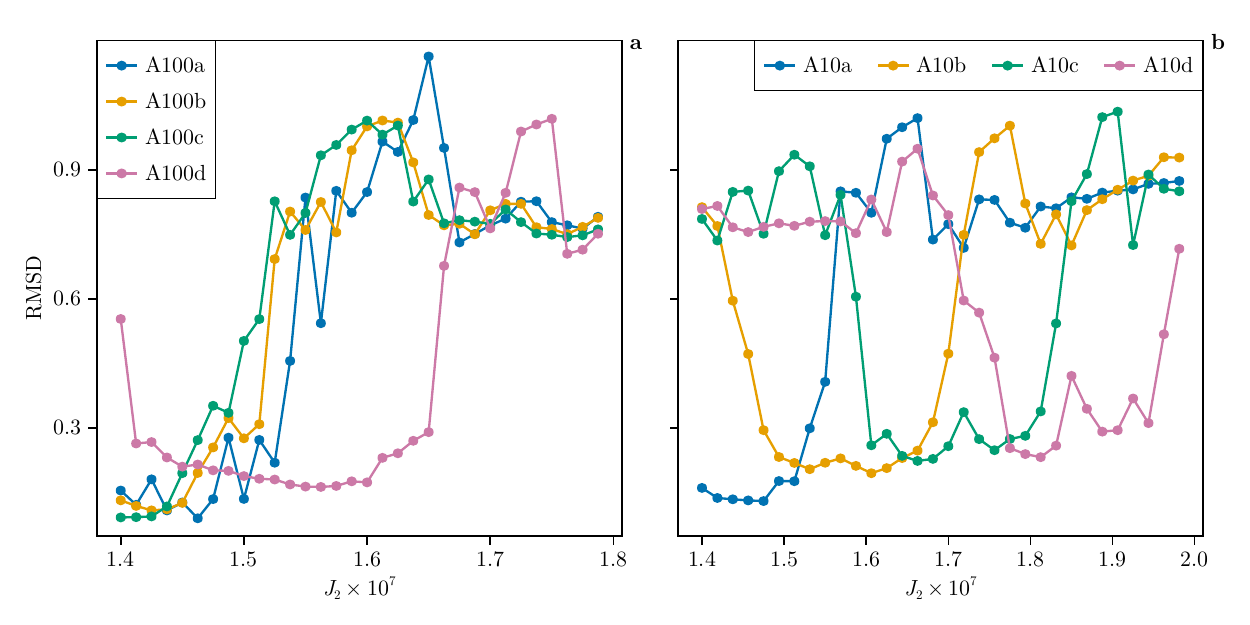}    
    \caption{
    \footnotesize
    \label{fig:rmsd}
    (a) RMSD scores contrasting the present solutions to \ZBETa\  
    between $-58$ and $-50$~Myr as a function of \JT\ values for four 
    groups of 100 asteroids (see text).
    (b) Same as (a), but for four groups of 10 asteroids with
    artificially increased asteroid mass by a factor of 
    $\fmA = 1.2, 1.4, 1.6,$ and $1.8$ (see text).
    }
\end{figure*}

\subsection{Numerical Results}

The numerical integrations that either resulted 
in the best or worst matches to the \ZBETa\ reference
solution showed similar patterns in Earth's eccentricity 
across all experiments (see Fig.~\ref{fig:ecc} for
illustration). The solutions with the lowest RMSD scores
matched the reference solution very well between $-58$ 
and $-50$~Myr, while those with the highest RMSD scores 
typically provided a poor match prior to
$-52$~Myr.
For the \texttt{A100a}-\texttt{c} experiments 
that include 100 asteroids, the solutions with lower \JT\
values (up to $1.525 \e{-7}$) most closely matched the 
\ZBETa\ reference solution (RMSD~$<0.25$) to $-58$~Myr 
(see Fig.~\ref{fig:rmsd}a).
However, inclusion of the 
100 most massive asteroids (\texttt{A100d}) allowed 
for a greater increase in \JT\ values (up to 
$\sm{1.65}\e{-7}$) before the solutions rapidly 
diverged from \ZBETa\ for larger \JT\ values
(Fig.~\ref{fig:rmsd}a). 
This result suggested that just increasing the
asteroid mass (instead of number) 
may allow for increased \JT\ values.
In the {\tt A10} simulations, we found that
for each increase in asteroid mass ({\tt a-d}),
a higher \JT\ value resulted in a good
match with the \ZBETa\ reference solution 
(Fig.~\ref{fig:rmsd}b).
For example, we found that by artificially increasing 
the mass of the 10 named asteroids 
by $\x \fmA = 1.8$ (\texttt{A10d}),
\JT\ values of $1.775-1.8125 \e{-7}$ resulted
in astronomical solutions practically
indistinguishable from \ZBETa\ to $-58$~Myr
(RMSD~$\simeq 0.233$).
While the {\tt A10a-c} groups also showed
small RMSDs for certain \JT\ values (even smaller
RMSDs, see Fig.~\ref{fig:rmsd}b), we highlight 
{\tt A10d} here because it allows the largest \JT\ 
value that is most compatible with modern estimates 
(see Section~\ref{sec:intro}). Also, note that
all solutions with RMSD~$\lesssim 0.25$ provide
very good matches with \ZBETa\ from $-58$ to 
$-50$~Myr (for illustration, see Fig.~\ref{fig:ecc}).

\section{Analysis: Identifying a Mechanism \label{sec:mech}}

Our long-term numerical integrations show that an increased 
$J_2$ and a larger asteroid population/mass tend to cancel 
each other out in the long run. The result is opposite to 
the low-order theoretical guess (Section~\ref{sec:guess})
that suggests a mutual enhancement of
\JT\ and asteroid mass. In the following, we present an
analysis of our numerical integration results and suggest 
a mechanism that reconciles the ostensibly opposing outcomes
obtained in Sections~\ref{sec:guess} and~\ref{sec:num}.

\subsection{Secular Frequencies and Associated Angles
\label{sec:angl}}

The time scale on which differences in the numerical solutions
due to \JT\ and \mA\ become discernible is $\gtrsim 40$~Myr, 
suggesting that changes in secular frequencies (changes 
relative to the solution \ZBETa) are involved.
In particular, \citet{zeebelourens19} identified a specific 
chaotic resonance transition in \ZBETa\ just prior to $-40$~Myr,
involving the secular terms \gftL\ and \sftL. Thus, we begin
with an analysis of the relevant secular frequencies and 
associated angles in the numerical solutions.

The angles (aka arguments) associated with the secular frequency
terms were determined following \citet{lithwick11} and 
\citet{zeebelantink24aj}. Consider the classic variables for each 
planet (also used for FFT analysis, 
Section~\ref{sec:winFFT}):
\begin{eqnarray}
h =  e \sin(\vpi)         \quad & ; & \quad
k =  e \cos(\vpi)         \label{eqn:hk} \\
p = \sin (I/2) \ \sin \Om \quad & ; & \quad
q = \sin (I/2) \ \cos \Om \label{eqn:pq} \ ,
\end{eqnarray}
where $e$, $I$, $\vpi$, and $\Om$ are eccentricity, 
inclination, longitude of perihelion, and longitude of
ascending node, respectively. Next use $\sin(I/2) \simeq I/2$ 
(applicable to small $I$, as in our solutions). The variable 
pairs $(h,k)$ and $2(p,q)$ can then be combined 
into two complex variables for each planet, 
because $e^{\i \theta} = \cos \tht + \i \sin \tht$,
where $\i = \sqrt{-1}$:
\begin{eqnarray}
z     & = & e \exp(\hat{\imath} \ \vpi) \\
\zeta & = & I \exp(\hat{\imath} \ \Om ) \ .
\end{eqnarray}
The $(z, \zeta)$ for Earth were determined 
from Earth's computed orbital elements. Next, we applied a simple 
bandpass filter (rectangular window) centered on the 
fundamental frequencies $g_i$ and $s_i$ of interest 
(index $i = 3,4$).
The bandpass filter uses FFT and inverse FFT, and sets all 
frequencies to zero, except those within a rectangular 
window given by the interval of center frequency $\pm{20}$\%.
Note that the frequencies \gt, \gf\ and \st, \sf,
are pairwise very similar (ca.\ 17.37, 17.92~\asy\ 
and $-$18.85, $-$17.75~\asy, i.e., within 3\% and 6\%
of each other,
respectively), see \citet{zeebe17aj}. A sufficiently 
wide bandpass filter therefore generally includes both $g$- 
(respectively $s$) frequencies for $i = 3,4$, which is also the 
case in our analysis.

\begin{figure*}[t]
\begin{center}
\vspace*{-35ex}
\includegraphics[scale=0.7]{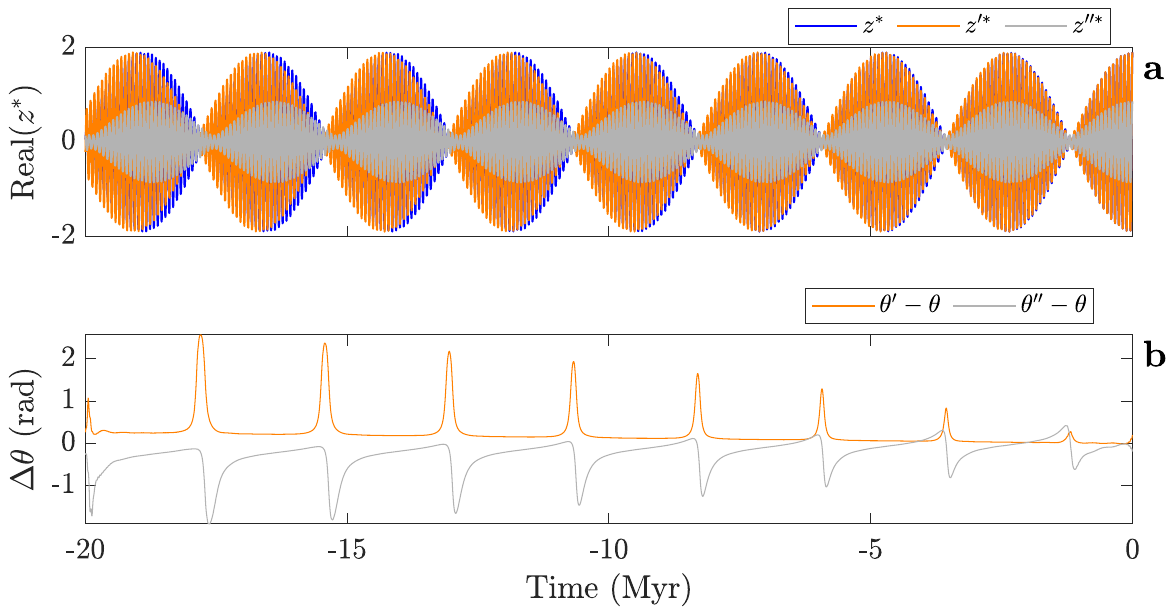}
\end{center}
\vspace*{-38ex}
\caption{
\footnotesize
Illustration of synthetic filter outputs \zs\ for \gt\
and \gf\ (see Eq.~(\ref{eqn:zstar})) and associated angles;
$(\gt, \gf) = (17.37, 17.92)$~\asy. The real part of the
complex \zs's is shown.
(a) Amplitudes and frequencies used in Eq.~(\ref{eqn:zstar}) 
for \zs\ (blue), \zps\ (orange), and \zpps\ (gray) are 
$a_3 = [1\ 1\ 0.5]$, $a_4 = [0.9\ 0.9\ 0.36]$ and
$\del_3 = $[0 1 $-$1] $\cdot \ 3\e{-4}$. 
The amplitude for \zpps\ (gray) was reduced for visual clarity.
(b) Associated differences in angles (aka arguments)
$\tht'-\tht$ (orange) and
$\tht''-\tht$ (gray).
For spikes in \Dtht\ occurring around the nodes, see 
Appendix~\ref{sec:frq}.
\label{fig:zzp}
}
\end{figure*}

Angles ($\tht$) were calculated from the filtered (complex) quantities 
$(z^*, \zeta^*)$ as $\vpi^* = \atant \{ \Im(z^*), \Re(z^*) \}$ 
\footnote{
$\atant(y,x)$ returns the four-quadrant inverse tangent 
of $y$ and $x$.}
and $\Om^*  = \atant \{ \Im(\zeta^*), \Re(\zeta^*) \}$,
where $\Im$ and $\Re$ denote the imaginary and real part
of a complex number. For illustration, a simple sinusoidal 
function $\exp(\hat{\imath} \ 2 \pi f t)$ with a single, 
constant frequency $f$ yields the angle
$\tht = 2 \pi f t$. Upon applying the filters for \gt, \gf\ 
and \st, \sf, the complex variables are approximately
given by:
\begin{eqnarray}
z^*   & \simeq & \ a_3 \exp(\hat{\imath} \ 2 \pi g_3 t) 
                 + a_4 \exp(\hat{\imath} \ 2 \pi g_4 t) \label{eqn:zstar} \\
\zt^* & \simeq & \ b_3 \exp(\hat{\imath} \ 2 \pi s_3 t) 
                 + b_4 \exp(\hat{\imath} \ 2 \pi s_4 t) \ ,
\end{eqnarray}
where $a_i$ and $b_i$ are (generally time-dependent) amplitudes.
In this case, the corresponding angles are not just 
associated with one, but two, secular
frequencies, which is of minor importance though because we
are interested in the {\sl difference} between angles
for slightly different sets of secular frequencies. For 
example, let $z^*$ depend on \gt\ and \gf\ and $z'^*$
on $g'_3 = \gt (1 + \del_3)$ and $g'_4 = \gf (1 + \del_4)$,
with $|\del_i| \ll 1$.
The superposition (interference) of $z^*$ and $z'^*$ then 
create characteristic patterns in the angles $\tht$ and 
$\tht'$ and hence in the difference $\tht'-\tht$ (for 
illustration with constant $a_i$ and $b_i$, see 
Fig.~\ref{fig:zzp}).

For example, a positive $\del_3$ increases \gt\ 
and hence reduces \gftL, which lengthens
the period $\gftL^{-1} \simeq 2.36$~Myr and
shifts the nodes of reduced amplitude that occur
about every 2.36~Myr further back in time (compare \zs\
and \zps, Fig.~\ref{fig:zzp}a).
Conversely, a negative $\del_3$ decreases \gt\ and raises 
\gftL, which shortens the period $\gftL^{-1}$
(compare \zs\ and \zpps, Fig.~\ref{fig:zzp}a).
Note that increasing
$\del_3$ or decreasing $\del_4$ by the same amount 
(and swapping the amplitudes) has virtually the same 
effect on \Dtht\ (except for flipping its sign). Importantly,
the pattern for \Dtht\ exhibiting spikes around the nodes
(superimposed on a long-term trend, Fig.~\ref{fig:zzp}b) 
is characteristic for differences in frequencies, not amplitudes.
Differences in amplitudes produce periodic patterns
in \Dtht\ alternating around $\Dtht = 0$ (not shown).
The cause for the spikes in \Dtht\ occurring around the 
nodes is discussed in Appendix~\ref{sec:frq}.

\subsubsection{Analysis of Solar System Integrations: 
A mechanism \label{sec:anyss}}

In the following, we apply the analysis described above to 
our full solar 
system integrations and examine whether or not characteristic 
patterns in the angles appear that are similar to those 
displayed in Fig.~\ref{fig:zzp}.

\begin{figure*}[t]
\begin{center}
\vspace*{-38ex}
\includegraphics[scale=0.7]{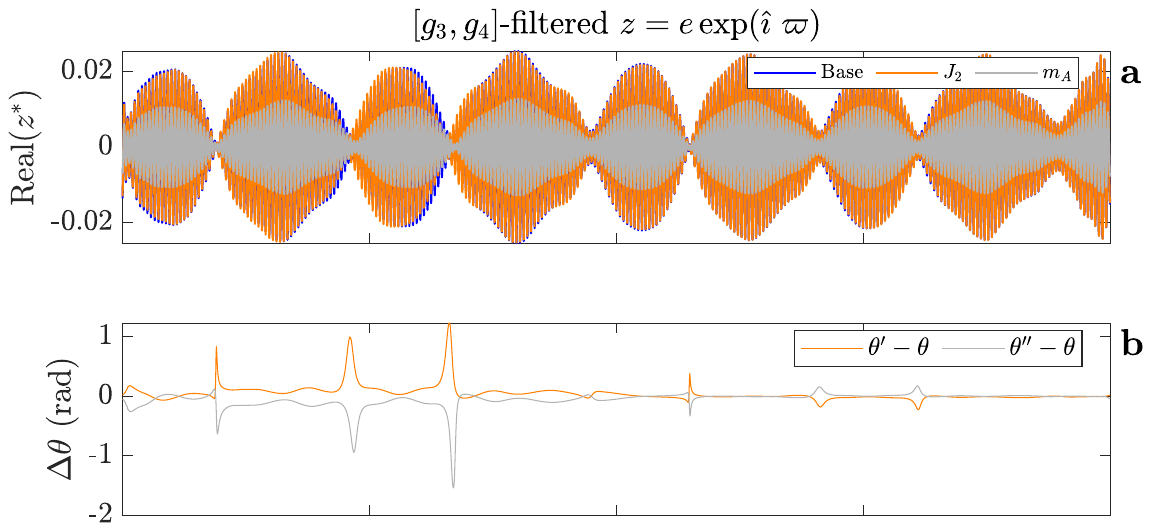}

\vspace*{-78ex}
\includegraphics[scale=0.7]{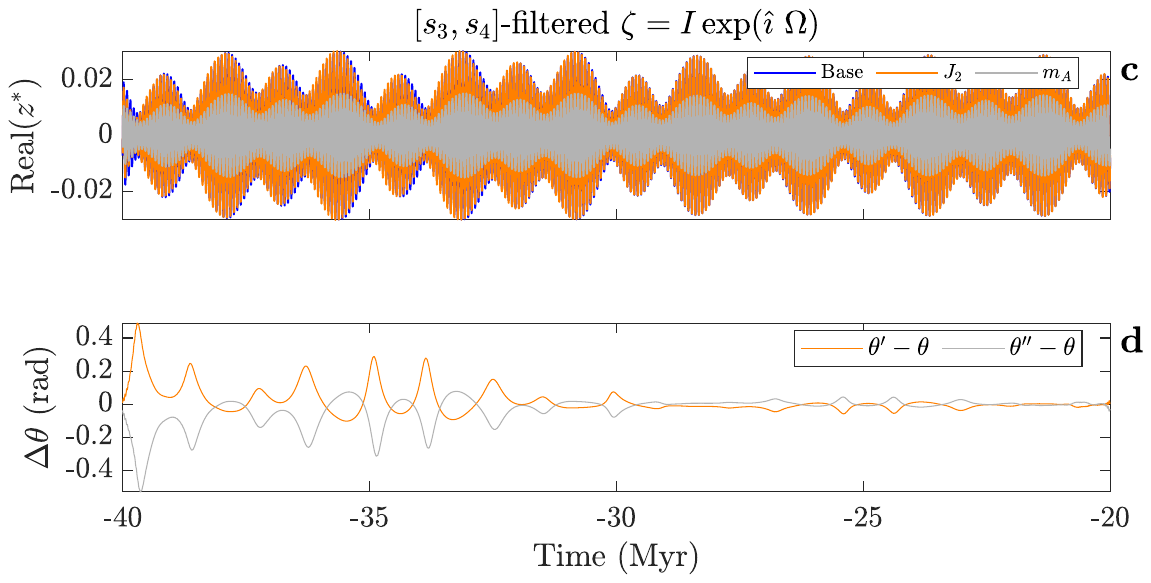}
\end{center}
\vspace*{-42ex}
\caption{
\footnotesize
Analysis of filter outputs \zs\ and \zts\ from full solar 
system integrations using the same approach as illustrated 
in Fig.~\ref{fig:zzp}.
Graphs labeled `\JT', and `\mA' show the effect 
of separate increases (relative to 'Base') in \JT\ and \mA. 
Runs used parameter
values (\JT,$\x \fmA$), with $\JT \x 10^7$ and $\fmA$ a factor 
describing the mass increase in \mA.
Base: (1.40,$\x 1.2$), 
\JT:  (1.98,$\x 1.2$), 
\mA:  (1.40,$\x 1.8$).
(a) Bandpass filter (rectangular window) applied to Earth's
$z = e \exp(\hat{\imath} \ \vpi)$
centered on the fundamental frequencies \gt\ and \gf\
(window: $\gt \pm 20\%$).
For plotting, the amplitude for `\mA' (gray) was reduced 
by 50\% for visual clarity.
(b) Associated differences in angles: 
$\tht'-\tht$ (\JT, orange) and
$\tht''-\tht$ (\mA, gray).
(c) Filter applied to Earth's
$\zeta = I \exp(\hat{\imath} \ \Om )$
centered on \st\ and \sf\ (window: $\st \pm 20\%$).
(d) Same as (b) but for \st\ and \sf.
Note that for the long-term evolution $-60 \lesssim t 
\lesssim -40$~Myr, the sign and larger trends in \Dtht\
older than about $-30$~Myr matter (not the small \Dtht\
changes in younger intervals).
\label{fig:omS}
}
\end{figure*}

\begin{figure*}[h]
\begin{center}
\vspace*{-35ex}
\includegraphics[scale=0.6]{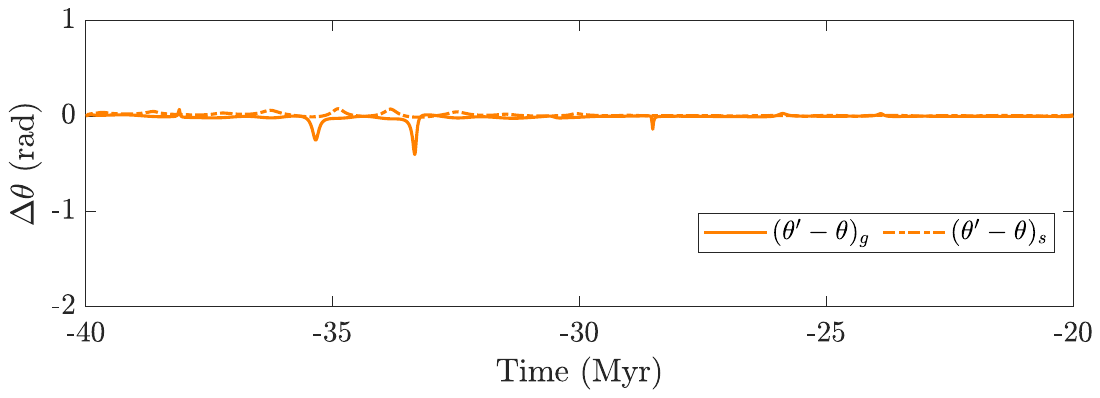}
\end{center}
\vspace*{-38ex}
\caption{
\footnotesize
Differences in angles \Dtht\ between a run with both \JT\ 
and \mA\ increased, i.e., (\JT,$\x \fmA$) = (1.8125,$\x 1.8$) 
and the Base-run (1.40,$\x 1.2$). 
Angles
$(\tht'-\tht)_g$ (solid orange) 
and
$(\tht'-\tht)_s$ (dashed orange)
are associated with $[\gt,\gf]$-filtered
$z = e \exp(\hat{\imath} \ \vpi)$
and $[\st,\sf]$-filtered
$\zeta = I \exp(\hat{\imath} \ \Om )$,
respectively. Note substantially reduced spikes 
in \Dtht\ compared to individually increased \JT\ 
and \mA\ (see Fig.~\ref{fig:omS}b,d).
\label{fig:doom}
}
\end{figure*}

We applied bandpass filters to Earth's 
$z = e \exp(\hat{\imath} \ \vpi)$ and
$\zeta = I \exp(\hat{\imath} \ \Om )$
centered on the fundamental frequencies (\gt, \gf)
and (\st, \sf), respectively, and calculated the
corresponding angles (see Fig.~\ref{fig:omS}).
Comparison of Fig.~\ref{fig:zzp} and~\ref{fig:omS} suggests
that increasing \JT\ tends to reduce \gftL\ going backward in 
time ($\lesssim -30$~Myr), while increasing \mA\ tends to 
raise \gftL\ in the long term (Fig.~\ref{fig:omS}a,b),
see discussion of ``differential effects'' below. 
Similarly, increasing \JT\ tends to reduce \sftA\ going 
backward in time ($\lesssim -30$~Myr), while increasing \mA\ 
tends to raise \sftA\ in the long term (Fig.~\ref{fig:omS}c,d).
As a result, for a specific set of \JT\ and \mA\ values, 
the two effects tend to cancel each other in long-term integrations.
Indeed, \Dtht\ between a run with both \JT\ and \mA\ increased, 
i.e., (\JT,$\x \fmA$) = (1.8125,$\x 1.8$) and the Base-run 
(1.40,$\x 1.2$), for instance, 
shows substantially reduced spikes and trends (see Fig.~\ref{fig:doom}), 
compared to individually increased \JT\ and \mA\ (Fig.~\ref{fig:omS}b,d).

Our analysis therefore suggests that ``differential effects'' on 
{\sl specific} secular frequencies involved in resonant terms
(i.e., \gftL\ and \sftL)
are critical in the long term, rather than short-term effects on the 
orbital elements of individual planetary orbits across the board (see 
Section~\ref{sec:guess}). Differential effect of \JT\ here 
means stronger/weaker effect on \gt\ vs.\ \gf, reducing \gftL, 
while differential effect of \mA\ means stronger/weaker effect 
on \gf\ vs.\ \gt, raising \gftL\ --- and correspondingly for the
$s$-frequencies. Notably, \JT's short-term effect is indeed 
larger for Earth's than for Mars' orbit and \mA's effect 
is larger for Mars' than for Earth's orbit (simply consider the
semimajor axes and relative distances to the Sun and asteroid,
respectively).
Mathematically, \JT's effect scales with $n/a^2$ 
(Eqs.~(\ref{eqn:omSJ}) and~(\ref{eqn:OmLJ})) and \mA's with 
$n(a/a_A)^2 \cdot \bo \propto n(a/a_A)^3$ 
(Eqs.~(\ref{eqn:OmLm}) and~(\ref{eqn:vpim})), where 
$\bo \propto (a/a_A)$ and $n \propto \sqrt{1/a^3}$. 
Importantly, 
however, there is generally no simple one-to-one relationship 
between a secular frequency and a single planet, 
particularly for the inner planets. The system's motion is 
a superposition of all eigenmodes, although some modes represent 
the single dominant term for some (mostly outer) planets
\citep[e.g.,][]{zeebe17aj,zeebekocken24esr}. 
Nevertheless, in the present case, the differential
effects that apply to the planetary orbits ($j = 3,4$)
also appear to apply to the secular frequencies ($i = 3,4$).
Finally, we will test whether or not the secular frequency shifts
inferred from \Dtht\ above can be confirmed using long-term 
spectral analysis (moving-window FFT, Section~\ref{sec:winFFT}).

\subsection{Moving-Window FFT \label{sec:winFFT}}

\begin{figure*}[t]
\begin{center}
\vspace*{-35ex}
\includegraphics[scale=0.7]{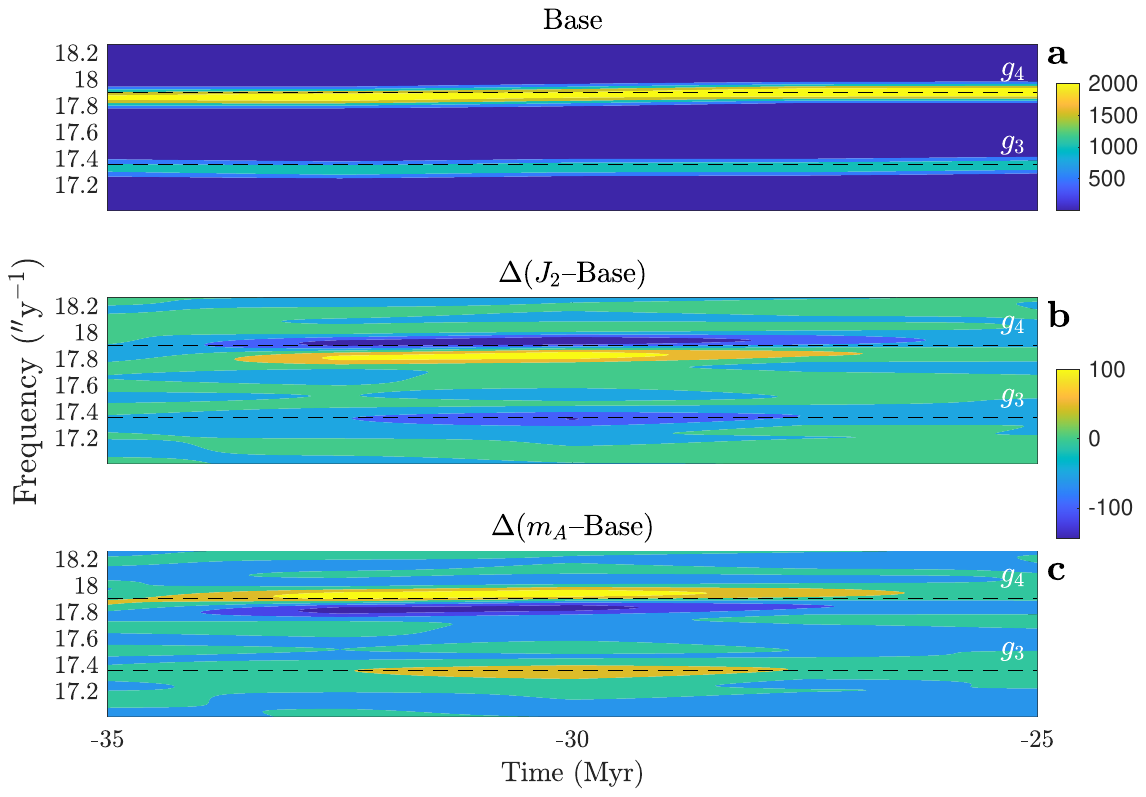}
\end{center}
\vspace*{-34ex}
\caption{
\footnotesize
Analysis of $g$ frequencies.
(a) Moving-window FFT power of Earth's $k =  e \cos(\vpi)$ 
of the Base-run (\JT,$\x \fmA$) = (1.40,$\x 1.2$)
using 10~Myr windows from $-40$ to $-20$~Myr in 2.5-Myr steps.
Dashed lines indicate \gt\ and \gf\ values based on 
FFT from $-20$ to $0$~Myr \citep[see][]{zeebe17aj}.
(b) Power difference between $k$'s moving-window FFT of 
increased \JT-run minus Base-run.
(c) Power difference between $k$'s moving-window FFT of 
increased \mA-run minus Base-run.
Note different power and colorbar scales in (a) vs.\ (b) and (c).
\label{fig:wfftg}
}
\end{figure*}

One valuable feature of our secular frequency analysis using
associated angles (Section~\ref{sec:angl}) is that it can
be applied over relatively short time scales (say $10^5$
to $10^6$~y). In contrast, secular frequency analyses
based on, e.g., Eqs.~(\ref{eqn:hk}) and~(\ref{eqn:pq}) using 
time-series analyses such as FFT usually require intervals
of tens of millions of years to separate and obtain accurate 
frequencies \citep[see e.g.,][]{zeebe17aj,zeebekocken24esr}.
For instance, we applied a moving-window FFT to Earth's 
$k =  e \cos(\vpi)$ using 10~Myr windows from $-40$ to 
$-20$~Myr in 2.5-Myr steps to follow the evolution of \gt\ 
and \gf\ in our Base-run with (\JT,$\x \fmA$) = (1.40,$\x 1.2$) 
(see Fig.~\ref{fig:wfftg}a). Based on an FFT analysis of a 
single solution, secular frequency changes are difficult
to identify for times younger than ca.\ $-40$~Myr because
the changes are small. However, 
trends in {\sl power differences} between solutions
at secular frequencies
are much easier to recognize (see Fig.~\ref{fig:wfftg}b,c).
For example, taking the difference between the moving-window
FFT of increased \JT\ minus Base (Fig.~\ref{fig:wfftg}b) 
shows reduced relative power along \gf\ (blue trough) and a 
shift toward lower frequency (yellow crest). However, note that 
the distance between the crest and trough in frequency space
(ca.\ $-0.1$~\asy) is not equal to the shift in \gf\ for 
$\D(\JT-$Base). Rather, it is indicative of the frequencies
with the largest rise and drop in relative power for 
$\D(\JT-$Base). The actual shift in \gf\ is much smaller 
(ca.\ $-0.005$~\asy). 

\begin{figure*}[t]
\begin{center}
\vspace*{-35ex}
\includegraphics[scale=0.7]{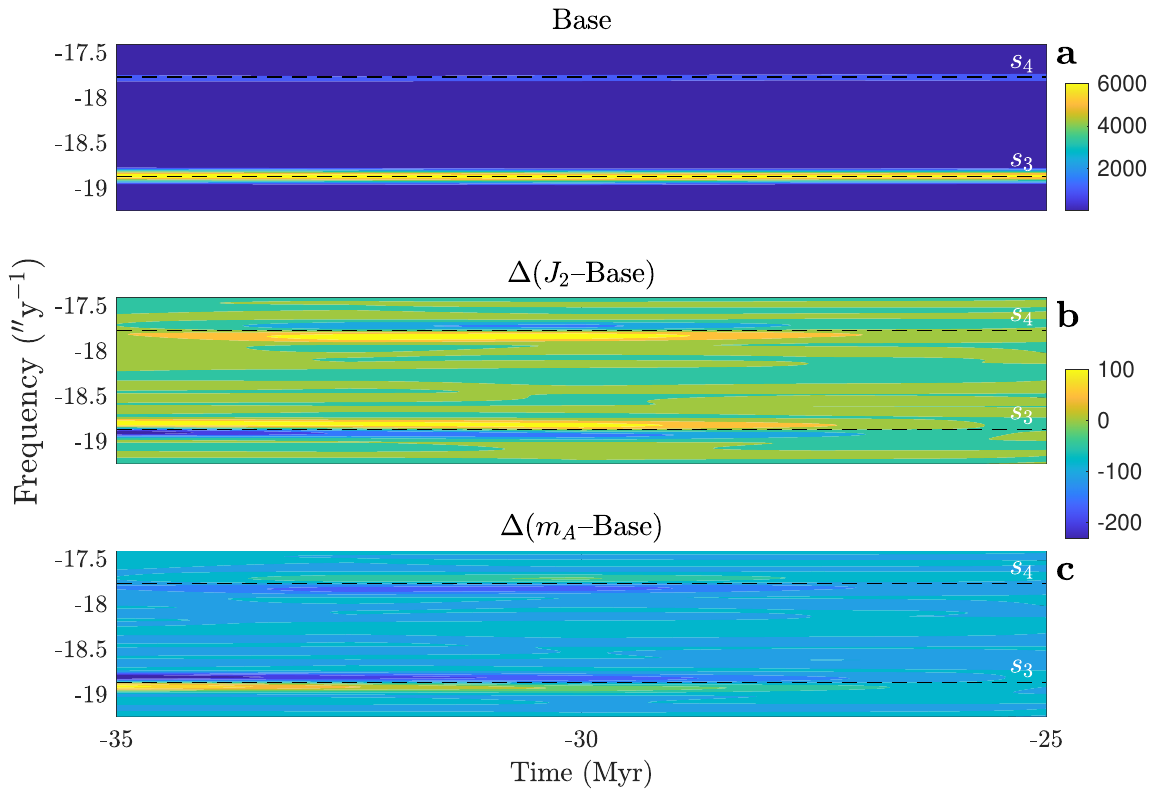}
\end{center}
\vspace*{-34ex}
\caption{
\footnotesize
Analysis of $s$ frequencies.
(a) Moving-window FFT power of Earth's $\hat{q} =  I \cos(\Om)$ 
of the Base-run (\JT,$\x \fmA$) = (1.40,$\x 1.2$)
using 10~Myr windows from $-40$ to $-20$~Myr in 2.5-Myr steps.
Dashed lines indicate \st\ and \sf\ values based on 
FFT from $-20$ to $0$~Myr \citep[see][]{zeebe17aj}.
(b) Power difference between $\hat{q}$'s moving-window FFT 
of increased \JT-run minus Base-run.
(c) Power difference between $\hat{q}$'s moving-window FFT 
of increased \mA-run minus Base-run.
Note different power and colorbar scales in (a) vs.\ (b) and (c).
\label{fig:wffts}
}
\end{figure*}

Nevertheless, the moving-window FFT confirms the secular 
frequency shift as inferred from \Dtht\ for the effect of
\JT\ (see Section~\ref{sec:angl}), i.e., a decrease 
in \gftL\ for $t \lesssim -30$~Myr.
Likewise, the moving-window FFT also confirms the 
secular frequency shift for the effect of \mA, i.e., an
increase in \gftL\ for $t \lesssim -30$~Myr.
Correspondingly, for the $s$-frequencies
the moving-window FFT confirms the decrease in \sftA\
due to \JT\ and the increase in \sftA\ due to \mA\ 
(Fig.~\ref{fig:wffts}).

Regarding secular frequency shifts, we
note that the premise of fundamental frequencies being 
solely a function of the masses and semimajor axes only holds 
for {\sl low-order} perturbation theory. In our full solar-system
integrations, the masses are constant and so are the 
semimajor axes, except for fluctuations around the mean
(no secular trends though, even on Gyr time scale, see e.g., 
\citet{zeebe15apjB}). Nonetheless, the integrations clearly
show shifts in fundamental frequencies, underlining the
solar system's strongly non-linear/chaotic behavior that 
is not captured by low-order perturbation theory.

\subsection{Recap: \JT-\mA\ Compensating Effect}
One main question of this study is, what 
ingredients are required for the \JT-\mA\ compensating 
effect to occur? Low-order perturbation theory on
short and intermediate time scales predicts the 
opposite outcome, i.e., reinforcing
effects of \JT\ and \mA\ (Section~\ref{sec:guess}).
The compensating effect is also absent
in a four-planet system (see Appendix~\ref{sec:four})
but is present over certain intervals in a six-planet 
system that includes
the inner planets, Jupiter, and Saturn (not shown).
The compensating effect is of course present in our 
full solar system integrations. The critical secular 
resonance discussed
above (Section~\ref{sec:anyss}), namely $m\gftL - \sftL$, 
where $m = 1,2$, is only present in more complex systems, 
i.e., in the six-planet system 
and the full solar system, confirming the resonance as an 
essential ingredient for the \JT-\mA\ compensating 
effect. Briefly, the mechanism identified here therefore
is the differential effect of \JT\ and \mA\ on the secular 
frequencies involved in the resonant terms \gftL\ and \sftL,
similar to the differential effect on Earth's and Mars' orbit
(Section~\ref{sec:anyss}). As mentioned in 
Section~\ref{sec:coc}, the mechanism is complex and
would not be captured even by a high-order perturbation 
theory including four-planet interactions 
(Appendix~\ref{sec:four}). It seems unlikely to us that, 
for instance, a high-order, six-planet perturbation theory 
would turn out to be much more instructive and reveal a 
simpler picture of the underlying mechanism than the one 
described here.

\section{Discussion and Conclusions}

In this contribution, we have demonstrated that a reduced solar 
quadrupole moment compensates for a diminished asteroid 
population (or total asteroid mass) in long-term solar system 
integrations over tens of millions of years. We have also 
presented an analysis of our numerical integration results 
that offers a mechanism for the long-term opposing effects of 
\JT\ and the number (or mass) of asteroids.
Our analysis suggests that differential effects on 
specific secular frequencies involved in resonant terms 
(i.e., \gftL\ and \sftL), are critical in the long term, 
rather than short-term effects on the orbital elements of 
individual planetary orbits across the board (see 
Section~\ref{sec:anyss}).

Our results should not be misinterpreted to suggest that,
in general, and regardless of the system and time scale, reducing 
the gravitational quadrupole moment of the dominant central 
mass in planetary systems can be used as a substitute for 
simulating dynamical asteroid contributions. Our results specifically 
apply to long-term integrations of the solar system 
($\mathcal{O}(10^8$~y)), with its particular structure of the 
inner planetary- and asteroidal orbits, and involving 
a specific secular resonance, namely $m\gftL - \sftL$, where
$m = 1,2$.

Regarding our knowledge of the current solar system, it is 
important that there are still notable uncertainties with 
respect to the accepted values of \JT\ and the total 
asteroidal mass. For example, recent studies (published 
over, say, the past 10-20 years) report modern \JT\ 
values of \sm{1.8} to $\sm{2.3}\e{-7}$ \citep[for summary, 
see e.g.,][]{rozelot20,zwaard22,alves25}. Recent estimates
for the total mass of the asteroid belt are 12.0 to $13.6\e{-10}\MS$
\citep[see e.g.,][]{kuchynkafolkner13,demeo13,pitjevapitjev18}.
For comparison, our integrations with 10 asteroids
and 80\% mass increase (total mass = $15.0\e{-10}\MS$),
yielded a solution practically indistinguishable 
from \ZBETa\ to $-58$~Myr using $J_2 \simeq 1.81 \e{-7}$.
Thus, the values used in our simulations are already close 
to the recent literature values 
cited above for the current solar system.
Furthermore, note that in our integrations with 100 
asteroids, only 10 were treated as HWPs and 90 as LWPs
(see Section~\ref{sec:num}). More realistic (but still 
too expensive) simulations would include $>$300 asteroids as 
HWPs, likely allowing further increases in \JT. We therefore conclude 
that if long-term intergrations including the full asteroid population 
were computationally feasible, a \JT\ value (within errors)
compatible with our current knowledge of the solar system 
could be supported.

Our findings lend confidence to our current solar system 
models, their numerical long-term integration, as well as 
the compatibility of our astronomical solutions with geologic 
data. Furthermore, because \JT\ and asteroids have a significant
impact on the long-term behavior of astronomical solutions,
our findings highlight that attempts to improve their accuracy 
by, e.g., tinkering with initial conditions using current/future 
astronomical observations are futile unless asteroid 
deficiencies in the underlying solar system model are
addressed.

\vspace*{5ex}

\begin{acknowledgments}
{\bf Acknowledgments.}
This research was supported by Heising-Simons Foundation Grant
\#2021-2800 and U.S. NSF grant OCE20-34660 to R.E.Z. 
We thank the reviewer for suggestions, which improved the manuscript.
\end{acknowledgments}

\software{
          \orbN, \giturl; on Zenodo: \zenurl, \cite{orbitn23}.
          \hnb\ {\tt v1.0.10}.
          }

\vspace*{5ex}

\appendix
\vspace*{-4ex}
\section{Basic integrator setup \label{sec:bset}}

Solar system integrations were performed following our earlier 
work \citep{zeebe15apjA,zeebe15apjB,zeebe17aj} with the integrator 
packages \orbN\ \citep{zeebe23aj} and \hnb\ \citep{rauch02} 
({\tt v1.0.10}) using the symplectic integrator and Jacobi 
coordinates \citep{wisdom91,zeebe15apjA}. For consistency with 
the original \ZBETa\ solution, we used \texttt{HNBody}
for the asteroid experiments (full solar system integrations
Set~1 and~2). The full simulations include contributions from 
general relativity, available as Post-Newtonian effects 
due to the dominant mass. The Earth-Moon system was modeled as a 
gravitational quadrupole \citep{quinn91} ({\tt lunar} option),
shown to be consistent with expensive Bulirsch-Stoer integrations 
up to 63~Ma \citep{zeebe17aj}. Initial conditions for the positions 
and velocities of the planets and Pluto for \ZBETa\ were generated 
from the DE431 ephemeris \citep{folkner14} using the SPICE toolkit 
for Matlab. The integrations for \ZBETa\ included 10 ``named''
asteroids (see Table~\ref{tab:ast}), with initial conditions generated 
at \url{ssd.jpl.nasa.gov/x/spk.html}. The 10~named asteroids
were treated as heavyweight particles, subject to the same full 
interactions as the planets. The unnamed asteroids (see 
Section~\ref{sec:nast}) were taken from 
a list of 300 asteroids ordered by semimajor axis as included in 
the DE430/431 ephimerides \citep{folkner14}.
Coordinates were obtained at JD2451545.0 
in the ECLIPJ2000 reference frame and subsequently rotated to 
account for the solar quadrupole moment ($J_2$) alignment with 
the solar rotation axis \citep{zeebe17aj}. Earth's orbital eccentricity 
and inclination from \ZBETa\ is available at \myurl\ and \npurl. 
Importantly, we provide results from $-100$ to 0~Myr but caution that 
the interval from $-100$ to ca.\ $-71$~Myr is unconstrained due to 
dynamical chaos \citep{zeebelourens19,zeebelourens22epsl,
kockenzeebe24pa}.

\section{Frequency Distortion at Nodes \label{sec:frq}}

\begin{figure*}[t]
\begin{center}
\vspace*{-30ex}
\includegraphics[scale=0.6]{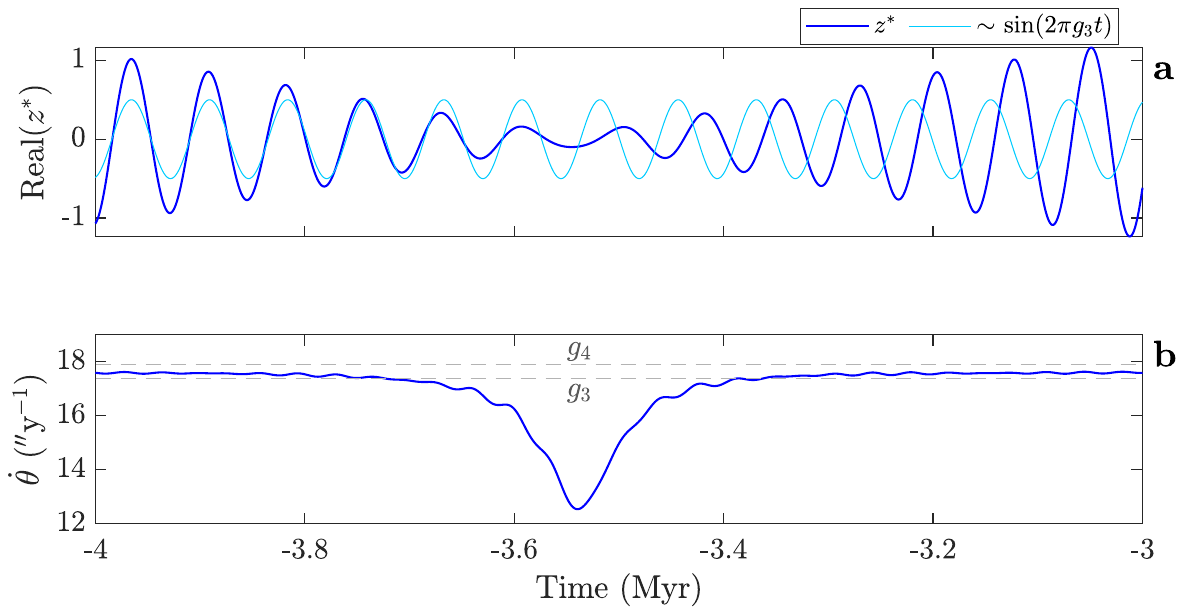}
\end{center}
\vspace*{-35ex}
\caption{
\footnotesize
Illustration of frequency distortion at nodes of reduced
amplitude for the synthetic \zs\ (see Eq.~(\ref{eqn:zstar}) 
and Fig.~\ref{fig:zzp}a).
(a) \zs\ and sinusoid with constant \gt\ frequency.
(b) Time derivative $\dot{\tht}$ of the angle \tht\ 
(divided by $2\pi$) associated with \zs.
\label{fig:frq}
}
\end{figure*}

As discussed in Section~\ref{sec:angl}, the \Dtht\ pattern
exhibiting spikes around the nodes is characteristic for 
differences in frequencies (Figs.~\ref{fig:zzp}b
and~\ref{fig:omS}b,d).
Here we describe the cause for the spikes in \Dtht\ occurring 
around the nodes. Zooming in on a node of the synthetic \zs\ 
(see Eq.~(\ref{eqn:zstar}) and Fig.~\ref{fig:zzp}a), 
illustrates the negative interference of the \gt\ and
\gf\ cycles that reduces the amplitude at the nodes
(Fig.~\ref{fig:frq}a). The time derivative $\dot{\tht}$
of the angle \tht\ (divided by $2\pi$) is a measure
of the frequency, which, except for the nodes, falls
between \gt\ and \gf\ (Fig.~\ref{fig:frq}b). At the nodes, 
however, the interference substantially distorts the 
cycles. For the example shown, the apparent period of \zs\ 
increases and the extrema
shift by almost half a cycle, compared to a 
sinusoid with constant \gt\ frequency (Fig.~\ref{fig:frq}a),
which substantially drops the frequency associated
with \zs's angle at the node (Fig.~\ref{fig:frq}b).
Importantly, the sign and magnitude of the frequency 
distortion strongly depends on the exact \gt\ and \gf\ 
values, as well as the amplitudes (see Eq.~(\ref{eqn:zstar})).
As a result, the differences in angles (\Dtht) between 
two solutions are amplified at the nodes, leading to
spikes in \Dtht\ as discussed in Section~\ref{sec:angl} 
(see Fig.~\ref{fig:zzp}b and~\ref{fig:omS}b,d).

\begin{figure*}[t]
\begin{center}
\vspace*{-30ex}
\includegraphics[scale=0.6]{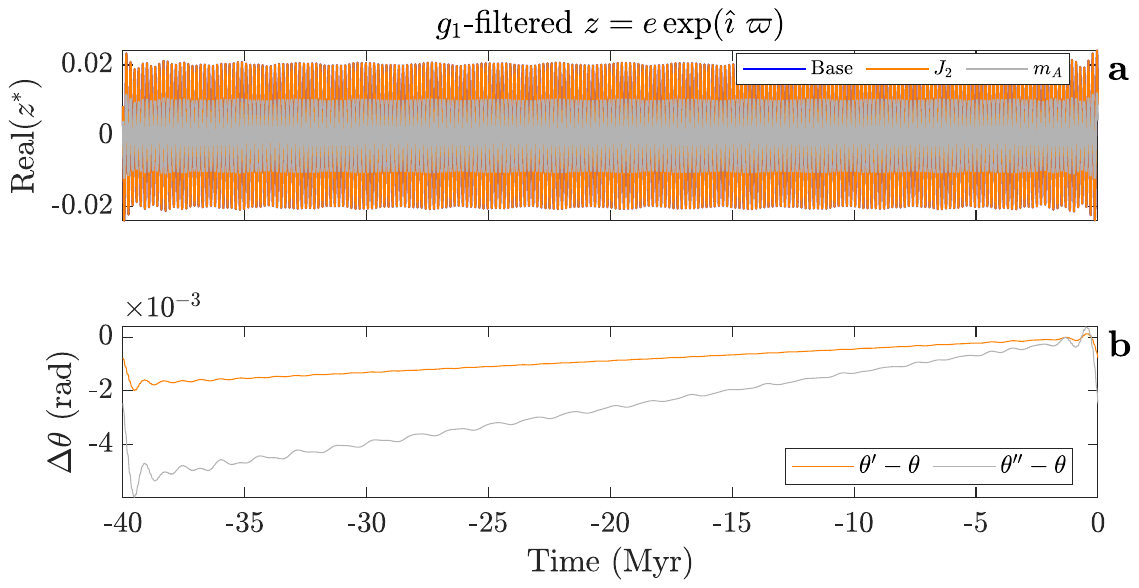}
\end{center}
\vspace*{-35ex}
\caption{
\footnotesize
Analysis of $g_1$-filter output for the four-planet
system using the same approach as illustrated 
in Fig.~\ref{fig:zzp}. 
(a) $g_1$-filter output for base, \JT, and asteroid runs.
Note: the blue line (base) is covered by the orange line (\JT),
i.e., the $g_1$ frequency shift from \JT\ is too small 
to be visible in this representation (same for `\mA').
For plotting, the amplitude for `\mA' (gray) was reduced 
by 50\% for visual clarity.
(b) Associated differences in angles (\Dtht): 
$\tht'-\tht$ (\JT, orange) and
$\tht''-\tht$ (\mA, gray).
The oscillations in \Dtht\ at the interval edges are 
artifacts due to filter distortion.
\label{fig:omSfp}
}
\end{figure*}

\section{Four-Planet System \label{sec:four}}
In the following, we show that the \JT-\mA\ compensating
effect is absent in a four-planet system. We include
Earth, Mars, Jupiter, and Saturn in the numerical
integrations with their standard masses and orbits,
which yields fundamental frequencies in the same 
ballpark as those of the full solar system for Venus, Earth, 
Jupiter, and Saturn \citep[see][]{zeebe17aj}. For example, 
the four $g$ frequencies are
$g_i = [7.651;\ 17.416;\  4.037;\ 27.995]$~\asy, 
corresponding to periods of 
$P_i = [169,398;\ 74,413;\ 321,033;\ 46,293]$~years.
Note that including fewer planets would yield some 
very long periods. Importantly, all four frequencies are 
well separated and the resonance dominated by Earth's and
Mars' orbit that is present in the full solar system is 
absent in the four-planet system. The system was integrated
covering the interval [$-$40 0]~Myr, including a base
run, \JT, and the asteroid Ceres (mass \mA). 
Subsequently, the same analysis was applied as 
described in Section~\ref{sec:angl}.

\begin{figure*}[t]
\begin{center}
\vspace*{-30ex}
\includegraphics[scale=0.6]{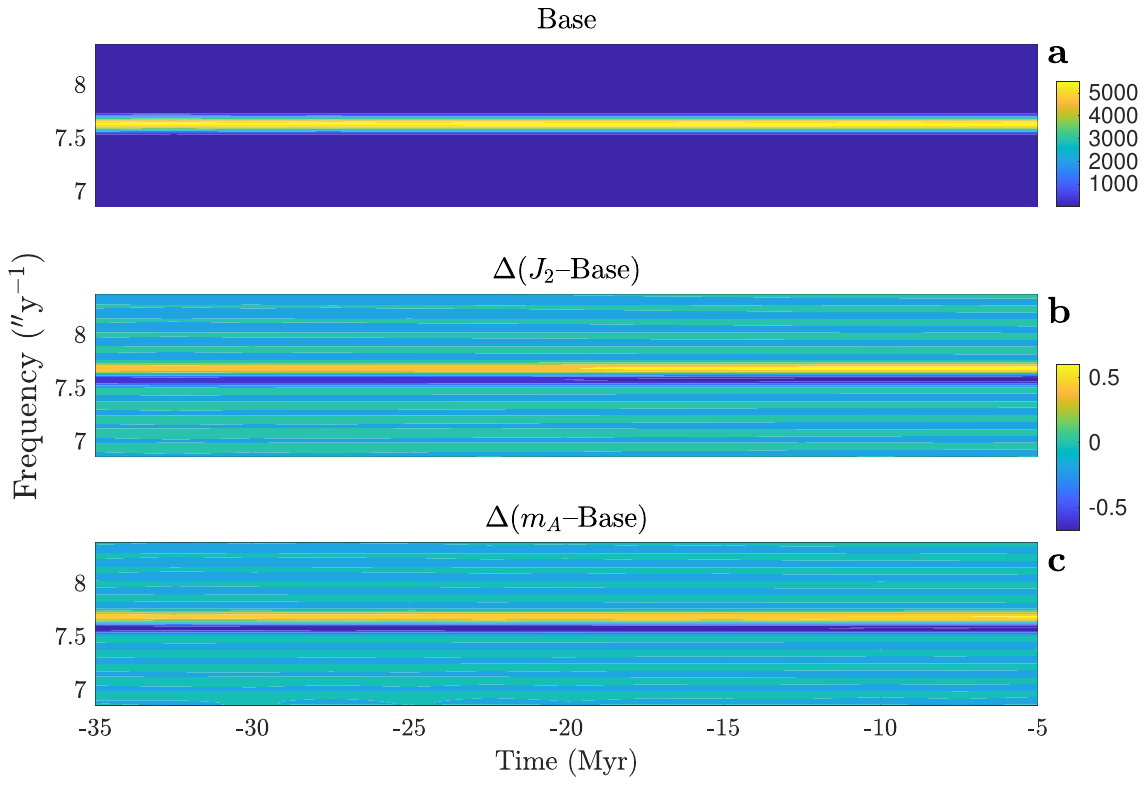}
\end{center}
\vspace*{-30ex}
\caption{
\footnotesize
Analysis of $g_1$ frequency for the four-planet system.
(a) Moving-window FFT power of Earth's $k =  e \cos(\vpi)$ 
of the base run using 10~Myr windows from $-40$ to $0$~Myr 
in 2.5-Myr steps.
(b) Power difference between $k$'s moving-window FFT of 
\JT-run minus base run.
(c) Power difference between $k$'s moving-window FFT of 
\mA-run minus base run.
Note different power and colorbar scales in (a) vs.\ (b) and (c).
\label{fig:wfftg-4p}
}
\end{figure*}

Integration of the four-planet system shows a regular
pattern for e.g., the $g_1$-filtered $z$'s and yields a linear
trend in \Dtht\ between the base- and \JT\ runs with an 
overall \Dtht\ of $\sm{2\e{-3}}$~rad over 40~Myr
(see Fig.~\ref{fig:omSfp}). Note that the oscillations in 
\Dtht\ at the interval edges are artifacts due to filter 
distortion. For comparison, $2\e{-3}$~rad over 40~Myr
corresponds to an average rate of 1~\masc\ $= 1\e{-5}$~\asy.
We also obtain a linear trend in \Dtht\ between the base 
and \mA\ run with an overall \Dtht\ of $\sm{5\e{-3}}$~rad over 
40~Myr (Fig.~\ref{fig:omSfp}b). The results show that \JT\
and \mA\ have reinforcing effects in the four-planet system
and that the \JT-\mA\ compensating effect (as in the full 
solar system) is absent. The moving-window FFT confirms
the result, indicating a $g_1$ increase for both the 
\JT\ and asteroid run (Fig.~\ref{fig:wfftg-4p}).
Again, note that the yellow and blue stripes in 
Fig.~\ref{fig:wfftg-4p}b and~c are indicative of the 
frequencies with the largest rise and drop in relative 
power (see colorbar), not the actual frequency shift.
Detecting a frequency shift of e.g., $1\e{-5}$~\asy\ 
($= 7.7\e{-9}$~kyr$^{-1}$) requires very high resolution.
Also note the fully ordered and regular pattern in 
power difference around $g_1$ for the four-planet system 
(Fig.~\ref{fig:wfftg-4p}b and~c), compared the chaotic pattern 
of the full system around $g_3$ and $g_4$ (Fig.~\ref{fig:wfftg}b 
and~c).

\clearpage

\bibliography{astJ2.bbl}{}
\bibliographystyle{aasjournal}

\end{document}